\documentclass[11pt,oneside,a4paper]{article}
\usepackage{hyperref}
\usepackage{natbib}
\usepackage{cancel}
\usepackage{graphicx}
\usepackage{amsmath}
\usepackage{systeme}
\title{\textbf{Spatially modulated plasma profile for turbulence and instabilities mitigation in fusion plasma.}}

\author{Ilya Shesterikov}

\begin{document}
	\maketitle
	
		Email: \href{mailto:ilyshes@gmail.com}{ilyshes@gmail.com}
	
	\begin{abstract}
		
This work explores a novel approach to mitigating turbulence in fusion plasmas through spatially modulated plasma profiles. By imposing a harmonic modulation on plasma parameters, we introduce conditions that alter the propagation characteristics of turbulent and MHD waves, a primary source of transport and instabilities in fusion devices. This modulation approach resembles bandgap formation in solid-state and photonic crystals, where spatial periodicity suppresses wave propagation within specific frequency bands. The mathematical framework developed here essentially resembles the parametric resonance of the harmonic oscillator. It reveals how a controlled spatial variation of turbulent wave phase velocity can effectively attenuate turbulence and instabilities. Several methods for implementing this modulation in plasma, including RF waves, static magnetic field perturbations, and modulated density profiles, are proposed as potential paths for achieving stable confinement. This concept could provide a versatile and potentially more controllable alternative to existing turbulence suppression techniques, with the goal of improving stability and confinement across a variety of magnetized fusion configurations.
		
	\end{abstract}

   The purpose of this work is to demonstrate the potential and new prospects that open up for fusion research through the creation of a spatially modulated plasma profile.
    These emerging opportunities can be used for suppressing instabilities and plasma turbulence (and actually plasma heating). In this work, we will focus primarily on using the proposed concept to suppress turbulence in plasma. Rather than providing a ready solution, this work aims to identify potential research directions and introduces a conceptual framework for mitigating plasma turbulence and instabilities.
   
   The formation of turbulence and plasma instabilities is one of the fundamental problems of nuclear fusion. Drift and interchange turbulence, electron-temperature and ion-temperature turbulence are the main causes of transport in fusion devices. Along with turbulence, there are also MHD instabilities that cause transport and loss of stability in toroidal devices. 
   Methods for suppressing or reducing turbulence already exist, such as

   \begin{itemize}
   	\item formation of transport barriers based on plasma rotation shear, leading to improved confinement modes
   	\item plasma shaping and magnetic field configuration optimization
   \end{itemize}
   
    However, the formation of transport barriers is a self-organizing process, not always well-controllable and manageable, and not implementable in all devices.
   Moreover, transport barriers are fairly localized in the radial direction—they do not suppress turbulence over a broad radial range. 
    Plasma shape and magnetic topology optimization is the subject of another constraints, as improper configurations could lead to new instability modes.
   
In other words, finding alternative approaches of turbulence and instabilities mitigation that are more universal, controllable, and comprehensive is very important. 
This is especially important considering the variety of magnetic confinement fusion devices that have emerged recently.

	Let's consider various types of devices that utilize magnetic fields to confine hot plasma, such as tokamaks, stellarators, pinches, and linear machines. 
	 
	The following sections will explore strategies for suppressing drift wave turbulence. While the specific details may vary, 
	the core principles of the proposed concept can be extended to other types of wave-like turbulence, such as interchange turbulence or MHD instabilities.
	
	 The time-averaged spatial scale of plasma parameters in fusion plasmas has typically the scale of the confinement device,
	  i.e. on the spatial scale of plasma turbulence  (which is typically $\approx 1-2~cm$ ) all plasma parameters appear uniform.
	 Consequently, parameters such as the phase velocity of waves, pressure gradient, and other quantities that influence the development of turbulence also appear uniform on these small turbulent scales.

	Let us now consider a scenario where the \textit{turbulent wave phase velocity} along the direction of turbulence propagation is spatially modulated but constant in time. In other words, we superimpose a harmonically varying profile with a wavelength comparable to the turbulence scale onto a smooth, homogeneous plasma profile. 
\textbf{	The presence of such a modulation significantly alters the nature of wave propagation and turbulence development.}
	 For instance, taking the drift wave as an example, the spatial modulation of the plasma density or magnetic field is equivalent to modulating the phase velocity of drift waves, as their phase velocity is described by the electron diamagnetic drift velocity.
	
	\subsection{Analogies to other physical systems}
	Suppressing waves by spatial modulation of their phase velocity is not a new phenomenon in general physics. This principle is similar to the existence of forbidden energy bands in the crystal lattice of a solid state [1]. Forbidden electron energy bands are formed due to the spatial periodicity of the potential energy field created by the  crystal lattice. Solving the Schrödinger equation for the electron wave function in a periodic potential field gives us the band structure of energy.
	
	Another example is optical photonic crystals [2,3,4] . The propagation of light in a medium with a spatially periodic refractive index leads to the formation of the so-called optical band structure, zones of forbidden and allowed EM wave frequencies that either can or cannot propagate in the crystal. This is also easily reproduced if we look at the solution of the wave equation in a medium with a spatially varying refractive index.
	
	In the field of mechanics, this phenomenon is commonly referred to as \textit{parametric instability} [5,6,7,8,9,10]. Depending on specific resonant conditions, it can cause mechanical oscillations to either amplify or diminish.

	\section{Mathematical formalism}

	Let us consider the mathematical formalism of this process. As a test-bed  in our discussion we choose drift waves, although on the place of drift waves can be any other type of plasma waves (interchange-type turbulence or MHD waves). 
	Without delving into details, we write the wave equation for drift waves. 
	
	 The phase velocity of these waves, as can be seen, is determined by the electron diamagnetic drift velocity, i.e., it is inversely proportional to the magnetic field strength and plasma density. Thus, a spatial periodic variation of these quantities is equivalent to a spatial variation of the phase velocity.

	The equation descibing the propagation of the drift wave in the magnetized plasma can be written as follows:
	
	\begin{equation} \label{drift_wave}
		\dfrac{d^2 \textbf{n}}{dt^2} = u_p^2 \dfrac{d^2 \textbf{n}}{dx^2}
	\end{equation}

	here $u_p$ is the wave phase velocity. The drift wave phase velocity  represent the electron diamagnetic drift velocity:
	
	\begin{equation*}
		u_p = \dfrac{\nabla p \times \textbf{B}}{n e \textbf{B}^2 },
	\end{equation*}
	 where $p = n T_e$ is the plasma pressure.

First, we take the time Fourier transform of both sides of the Eq.\ref{drift_wave}.
	
\begin{equation}
		\mathcal{F} \left\{ \frac{\partial^2 \mathbf{n}}{\partial x^2} \right\} = \frac{\partial^2 \textbf{n}_{\omega}(x)}{\partial x^2}
\end{equation}	

\begin{equation*}
	\mathcal{F} \left\{     \dfrac{d^2 \textbf{n}}{dt^2}         \right\} = -\omega^2 \textbf{n}_{\omega}
\end{equation*}

	Combining both parts we rewrite the equation above in the Fourier space.

\begin{equation}\label{drift_wave_fourier}
u_p^2 \frac{\partial^2 \textbf{n}_{\omega}(x)}{\partial x^2} = 	-\omega^2 \textbf{n}_{\omega}
\end{equation}

	Let us rewrite this equation as a traditional  equation for a harmonic oscillator.
	
\begin{equation}
	\label{eqn:harmoscill}
	 \frac{\partial^2 \textbf{n}_{\omega}(x)}{\partial x^2} = 	-\dfrac{\omega^2}{u_p^2} \textbf{n}_{\omega} = -k^2 \, \textbf{n}_{\omega}
\end{equation}

The drift wave favorably develops in the conditions where $u_p$ is constant on the wavelength scale.

Let's now consider the opposite situation, the development in the plasma where $u_p$ has a periodical spatial dependence.

Specifically, we will determine the conditions for wave development in a case where $u_p$ slightly differs from some constant value and is a simple spatially periodic function $u_p(x)$.

\begin{equation}\label{eqn:phase_velocity_modulation}
	u_p(x) = u^0_p (1 - \varepsilon \, \cos(\beta x))
\end{equation}	
	
where the constant $\varepsilon \ll 1$ and $\beta$ designates the wave number of the spatial modulation.
The sign of $\varepsilon$ is not that important since we can always change this sign by the corresponding choice of the reference frame.	
Substituting this expression (\ref{eqn:phase_velocity_modulation}) in the Eq.\ref{eqn:harmoscill}	gives
	
	\begin{equation}
		\label{eqn:wave_eq_1}
		\frac{\partial^2 \textbf{n}_{\omega}(x)}{\partial x^2} 
		= 	-\dfrac{\omega^2}{{  (u^0_p (1 - \varepsilon \, \cos(\beta x)) )   }^2} \textbf{n}_{\omega} \approx
		 -\bigl(\dfrac{\omega}{u^0_p }\bigl)^2 (1 + 2 \, \varepsilon \, \cos(\beta x)) \, \textbf{n}_{\omega}
	\end{equation}

or, designating 

\begin{equation}
	\label{eqn:k0}
k_0 = \dfrac{\omega}{u^0_p }
\end{equation}

\begin{equation}
	\label{eqn:wave_eq_2}
	\frac{\partial^2 \textbf{n}_{\omega}(x)}{\partial x^2} 
	= 	- k_0^2 (1 + 2 \, \varepsilon \, \cos(\beta x)) \, \textbf{n}_{\omega}
\end{equation}

We will see later that the effect of modulation is strongest if 
the  wavenumber $\beta$ is close to the  doubled wavenumber of drift wave $k_0$.
Therefore we will assume

\begin{equation}
	\label{eqn:betta}
\beta = 2 \, k_0 + \delta
\end{equation}

where $\delta$ is a small deviation of $\beta$ from $2 \, k_0$.

\begin{equation}
	\label{eqn:wave_eq_3}
	\frac{\partial^2 \textbf{n}_{\omega}(x)}{\partial x^2} 
	= 	- k_0^2 (1 + 2 \, \varepsilon \, \cos((2 \, k_0 + \delta) x)) \, \textbf{n}_{\omega}
\end{equation}

For the simplification we introduce the new designation 
 
\begin{equation}
	\label{eqn:epsilon}
	\epsilon = 2 \, \varepsilon
\end{equation}	
	
and rewrite the equation
\begin{equation}
	\label{eqn:Mathieu_eq}
	\frac{\partial^2 \textbf{n}_{\omega}(x)}{\partial x^2} 
	= 	- k_0^2 (1 + \epsilon \, \cos((2 \, k_0 + \delta) x)) \, \textbf{n}_{\omega}
\end{equation}	
	
The equations of this type are called in mathematics the \textbf{Mathieu equation}.	

Using the \textit{method of variation of parameters}, the solution   $\textbf{n}_{\omega}(x)$ to our transformed equation may be written as

\begin{equation}
	\label{eqn:solution_form}
	\textbf{n}_{\omega}(x) = a(x) \cos((k_0 + \frac{\delta}{2})x) + b(x) \sin((k_0 + \frac{\delta}{2})x)
\end{equation}	

where the rapidly varying components, $\cos((k_0 + \frac{\delta}{2})x)$ and   $\sin((k_0 + \frac{\delta}{2})x)$
 have been factored out to isolate the slowly varying amplitudes   $a(x)$ and   b(x).

Where (a(x)) and (b(x)) are slowly (compared to the factors cos and sin) changing functions of time. Such a solution, of course, is not exact. 
We proceed by substituting this form of the solution (\ref{eqn:solution_form})  into the differential equation (\ref{eqn:Mathieu_eq}) and considering that both the coefficients in front of   $\cos((k_0 + \frac{\delta}{2})x)$ and   $\sin((k_0 + \frac{\delta}{2})x)$ must be zero to satisfy the differential equation identically. We also omit the second derivatives of   $a(x)$ and   
$b(x)$ on the grounds that   $a(x)$ and   $b(x)$ are slowly varying functions.

\begin{gather}
	\label{eqn:system_1}
	  -2 \dfrac{da(x)}{dx} (k_0 + \frac{\delta}{2}) - b(x) (k_0 + \frac{\delta}{2})^2 + k_0^2 b(x) \big( 1 - \dfrac{\epsilon}{2}\big)  = 0 \notag\\
	    2 \dfrac{db(x)}{dx} (k_0 + \frac{\delta}{2}) - a(x) (k_0 + \frac{\delta}{2})^2  + k_0^2 a(x) \big( 1 + \dfrac{\epsilon}{2}\big)  = 0
\end{gather}

A more detailed derivation is provided in the Appendix.
Neglecting all terms above the first order in $\delta$, we can simplify

\begin{gather}
	\label{eqn:system_2}
	2 \dfrac{da(x)}{dx}    + b(x)  \delta     + \dfrac{k_0 b(x) \epsilon}{2}  = 0 \notag\\
	2 \dfrac{db(x)}{dx}    - a(x)  \delta     + \dfrac{k_0 a(x) \epsilon}{2} = 0
\end{gather}

We got the system of two first-order linear differential equations. 
To find the general solution of a system, the system can be expressed in matrix form as:

\begin{gather}
	 \dfrac{d\textbf{Y}}{dx} = A \textbf{Y},
\end{gather}

where 

\begin{gather}
	\textbf{Y} =
\begin{bmatrix}
	a(x) \\
	b(x)
\end{bmatrix}
=
c_1 \, \vec{V}_1 \, e^{\lambda_1 x} + c_2 \, \vec{V}_2 \, e^{\lambda_2 x},
\end{gather}

$\lambda_1$ and $\lambda_2$ are the eigenvalues of the matrix A

\begin{figure}[h]
	\centering
	\includegraphics[width=0.5\textwidth]{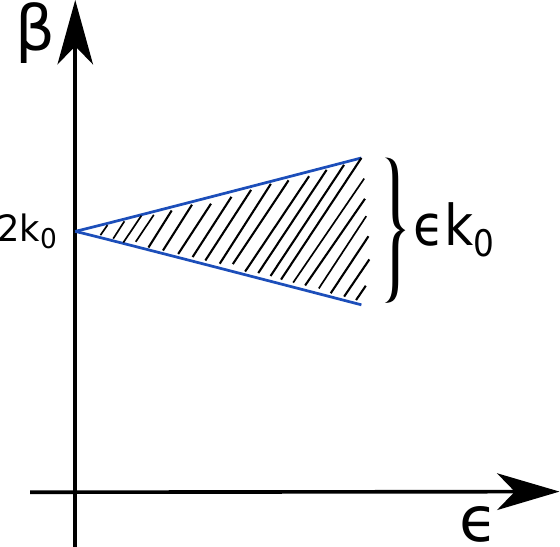}
	\caption{The chart of the parametric decay of a turbulent/instability waves.
		The effective damping occurs in the range of $\epsilon k_0$ around the $\beta$-wavenumber $2 k_0$.  
	}
	\label{fig:stability}
\end{figure}

\begin{gather}
	A = 
\frac{1}{2}
\begin{bmatrix}
	0 & \delta + \dfrac{\epsilon \, k_0}{2} \\
	-\delta + \dfrac{\epsilon \, k_0}{2} & 0
\end{bmatrix},
\end{gather}

\(\vec{V}_1\) and \(\vec{V}_2\) are corresponding eigenvectors, and \(c_1\) and \(c_2\) are some constants.

The eigenvalues are given by the expression

\begin{gather}
	\lambda_{1,2} = \pm \dfrac{1}{2}\sqrt{ \Big( \dfrac{\epsilon \, k_0}{2} \Big)^2 - \delta^2 }
\end{gather}

The condition for the occurrence of a wave attenuation is that $\lambda$ is real (i.e., \( \lambda^2 \geq 0 \)).
The parameter $\lambda$ characterizes the spatial attenuation (or amplification) of the wave.
Thus, it occurs in the interval of 

\begin{gather}
		\label{eqn:stability}
-\dfrac{\epsilon \, k_0}{2} < \delta < \dfrac{\epsilon \, k_0}{2}
\end{gather}
around the wavenumber $2 k_0$ of the $u_p(x)$ spatial modulation.

The chart in Figure (\ref{fig:stability}) presents the graphical illustration of the mathematical results above. 
The effective damping of turbulent or instability waves occurs within a range defined by $\epsilon \, k_0$ around the spatial modulation wavenumber $\beta=2 \, k_0$.

\section{Practical Significance}

\subsection{\textit{Modulation approach }}
The obtained result indicates that by creating a spatially modulated phase velocity profile, one establishes conditions to suppress drift waves. 
The main issue lies in finding a rational and feasible method for a modulation of plasma parameters. 
Here are some technical ways to implement this approach:

\begin{itemize}{}
	\item RF electromagnetic waves (Alfvén waves), which lead to perturbations in the plasma's magnetic field and, 
	therefore, drift wave phase velocity $u_p$.
  \item Amplitude modulation of the microwave electromagnetic waves, which lead to perturbations in the plasma density due to ponderomotive force and, 
	therefore, drift wave phase velocity $u_p$.
	
	  \item Externally driving an another plasma instability leading to the perturbation of the plasma magnetic field or density.
	
	\item A static magnetic field perturbation created by external currents.
	\item A spatially-modulated neutral particle beam.
\end{itemize}

In this paper we are not aiming to discuss the technical details on the implementation of these approach, this is the subject of separate works.

\subsection{\textit{Amplification vs. damping}}

As seen in the Equation \ref{eqn:stability}, the parameter $\lambda$ can be either positive or negative. From a purely mathematical perspective, this implies that the observed resonance can result in either amplification or decay of the propagating wave.

Whether the propagating wave instability is amplified or damped is a complex question that depends on numerous factors and the specific plasma modulation approach.
The amplification or damping of the wave is determined by the specific physical mechanisms that facilitate or inhibit energy and momentum exchange between the plasma instability wave and the imposed modulation.

For instance, the steady state static magnetic field $\textbf{B}$ spatial modulation created by an external currents (if this is feasible at all) will lead to the dumping since it is not the subject of the energy exchange with plasma waves. 
The low-frequency plasma modulation created by an externally launched wave have a more complex wave-wave interaction physics and can lead to both
 amplification and dumping, depending on the features of dispersion relations of both imposed modulation (on the one hand) and plasma instability waves (on the other hand).

Summarizing the results obtained above, it can be said that waves propagating in such a medium will experience attenuation. This result is well-known in many areas of physics that deal with oscillations or waves, and is therefore quite predictable.

\section{Conclusion}

This work has demonstrated that introducing a spatially modulated phase velocity profile in plasma holds promising potential for mitigating turbulence, specifically drift wave instabilities, in fusion plasma environments. The framework established here illustrates that spatial modulation creates conditions akin to bandgaps in solid-state physics, where specific frequencies are inhibited, thus attenuating wave propagation. This principle could serve as an alternative approach to current turbulence suppression methods, which frequently encounter practical limitations and challenges in implementation.

The practical feasibility of applying spatial modulation was also explored, offering several methods for implementing these profiles in plasma, such as using RF waves, modulated microwave electromagnetic waves, or static magnetic perturbations. These methods provide a foundation for further investigation into the optimal means of achieving effective modulation in various fusion device configurations. The study underscores the dual potential of modulation, highlighting that careful tuning of the parameters can either amplify or dampen wave instabilities, depending on specific interactions between imposed modulation and instability waves.
Future work should focus on feasibility study of these specific approaches, assessing their practicality, and testing their effectiveness in experimental settings to advance toward stable, high-performance fusion plasmas.

\section*{Bibliography}

\begin{enumerate}
		\item Kittel, C. (2005). Introduction to Solid State Physics. Wiley, Hoboken, NJ, 8th edition.
			\item Joannopoulos, J. D., Johnson, S. G., Winn, J. N., \& Meade, R. D. (2008). Photonic Crystals: Molding the Flow of Light. Princeton University Press,  
		Princeton, NJ, 2nd edition.
		
		\item Sakoda, K. (2001). Optical Properties of Photonic Crystals. Springer, Berlin, Heidelberg.
		
			\item Haus, J. W. (2016). Fundamentals and Applications of Nanophotonics. Woodhead Publishing, Cambridge.
	\item Strogatz, S. H. (2000). Nonlinear Dynamics and Chaos: With Applications to Physics, Biology, Chemistry, and Engineering. Westview Press.  
	\item Timoshenko, S., Young, D. H., \& Weaver, W. (1974). Vibration Problems in Engineering. Wiley.
		\item Den Hartog, J. P. (1947). Mechanical Vibrations. McGraw-Hill.
		\item Guckenheimer, J., \& Holmes, P. (1983). Nonlinear Oscillations, Dynamical Systems, and Bifurcations of Vector Fields. Springer-Verlag.  
		\item Thomson, W. T., \& Dahleh, M. D. (1998). Theory of Vibration with Applications. Prentice Hall.
		\item Svelto, O. (1998). Principles of Lasers. Springer.

\end{enumerate}

\section*{Appendix}

substituting this form of the solution (\ref{eqn:solution_form})  into the differential equation (\ref{eqn:Mathieu_eq}) and considering that both the coefficients in front of   $\cos((k_0 + \frac{\delta}{2})x)$ and   $\sin((k_0 + \frac{\delta}{2})x)$ must be zero to satisfy the differential equation identically. We also omit the second derivatives of   $a(x)$ and

Substituting the solution of the form (\ref{eqn:solution_form}) into the differential equation (\ref{eqn:Mathieu_eq}) gives

\begin{gather}
	\dfrac{d \textbf{n}_{\omega}(x)}{dx} = \dfrac{da(x)}{dx} \cos((k_0 + \frac{\delta}{2})x) - \notag\\
	a(x) (k_0 + \frac{\delta}{2}) \sin((k_0 + \frac{\delta}{2})x) + \notag\\
	\dfrac{db(x)}{dx} \sin((k_0 + \frac{\delta}{2})x) +   \notag\\
	b(x) (k_0 + \frac{\delta}{2}) \cos((k_0 + \frac{\delta}{2})x)
\end{gather}

\begin{gather}
	\dfrac{d^2 \textbf{n}_{\omega}(x)}{dx^2} = \xcancel{\dfrac{d^2a(x)}{dx^2} \cos((k_0 + \frac{\delta}{2})x)} - \notag\\
	\dfrac{da(x)}{dx} (k_0 + \frac{\delta}{2}) \sin((k_0 + \frac{\delta}{2})x) - \notag\\
	\notag\\
	\dfrac{da(x)}{dx} (k_0 + \frac{\delta}{2}) \sin((k_0 + \frac{\delta}{2})x) - \notag\\
	a(x) (k_0 + \frac{\delta}{2})^2 \cos((k_0 + \frac{\delta}{2})x) + \notag\\
	 \notag\\
	\xcancel{\dfrac{d^2b(x)}{dx^2} \sin((k_0 + \frac{\delta}{2})x)} +   \notag\\
	\dfrac{db(x)}{dx} (k_0 + \frac{\delta}{2}) \cos((k_0 + \frac{\delta}{2})x) +   \notag\\
	\notag\\
	\dfrac{db(x)}{dx} (k_0 + \frac{\delta}{2}) \cos((k_0 + \frac{\delta}{2})x) -   \notag\\
	b(x) (k_0 + \frac{\delta}{2})^2 \sin((k_0 + \frac{\delta}{2})x)
\end{gather}
We omit the second derivatives of   $a(x)$ and $b(x)$ and rewrite the last equation.

\begin{gather}
	\dfrac{d^2 \textbf{n}_{\omega}(x)}{dx^2} = 	-\dfrac{da(x)}{dx} (k_0 + \frac{\delta}{2}) \sin((k_0 + \frac{\delta}{2})x) - \notag\\
	 \notag\\
	\dfrac{da(x)}{dx} (k_0 + \frac{\delta}{2}) \sin((k_0 + \frac{\delta}{2})x) - \notag\\
	a(x) (k_0 + \frac{\delta}{2})^2 \cos((k_0 + \frac{\delta}{2})x) + \notag\\
	 \notag\\
	\dfrac{db(x)}{dx} (k_0 + \frac{\delta}{2}) \cos((k_0 + \frac{\delta}{2})x) +   \notag\\
	 \notag\\
	\dfrac{db(x)}{dx} (k_0 + \frac{\delta}{2}) \cos((k_0 + \frac{\delta}{2})x) -   \notag\\
	b(x) (k_0 + \frac{\delta}{2})^2 \sin((k_0 + \frac{\delta}{2})x) 
\end{gather}

Combining similar terms together we get

\begin{gather}
	\dfrac{d^2 \textbf{n}_{\omega}(x)}{dx^2} = 	-2 \dfrac{da(x)}{dx} (k_0 + \frac{\delta}{2}) \sin((k_0 + \frac{\delta}{2})x) - \notag\\
	 \notag\\	
	a(x) (k_0 + \frac{\delta}{2})^2 \cos((k_0 + \frac{\delta}{2})x) + \notag\\
	 \notag\\
	2 \dfrac{db(x)}{dx} (k_0 + \frac{\delta}{2}) \cos((k_0 + \frac{\delta}{2})x) -   \notag\\
	 \notag\\
	b(x) (k_0 + \frac{\delta}{2})^2 \sin((k_0 + \frac{\delta}{2})x) 
\end{gather}

\begin{gather}
	\dfrac{d^2 \textbf{n}_{\omega}(x)}{dx^2} = 	\big(  -2 \dfrac{da(x)}{dx} (k_0 + \frac{\delta}{2}) - b(x) (k_0 + \frac{\delta}{2})^2 \big) \sin((k_0 + \frac{\delta}{2})x) - \notag\\
	..... \notag\\	
	\big(  2 \dfrac{db(x)}{dx} (k_0 + \frac{\delta}{2}) - a(x) (k_0 + \frac{\delta}{2})^2 \big) \cos((k_0 + \frac{\delta}{2})x) 
\end{gather}

Now, we turn to simplify the RHS of the equation (\ref{eqn:Mathieu_eq}).

\begin{gather}
	(1 + \epsilon \, \cos((2 \, k_0 + \delta) x)) \, \textbf{n}_{\omega} \notag\\
	= \notag\\
	(1 + \epsilon \, \cos((2 \, k_0 + \delta) x)) \, \big(a(x) \cos((k_0 + \frac{\delta}{2})x) + b(x) \sin((k_0 + \frac{\delta}{2})x)	\big) 
\end{gather}

\begin{gather}
	(1 + \epsilon \, \cos((2 \, k_0 + \delta) x)) \, \textbf{n}_{\omega} \notag\\
	= \notag\\
	(1 + \epsilon \, \cos((2 \, k_0 + \delta) x)) \, \big(a(x) \cos((k_0 + \frac{\delta}{2})x) + b(x) \sin((k_0 + \frac{\delta}{2})x)	\big) 
	= \notag\\
	a(x) \cos((k_0 + \frac{\delta}{2})x) + b(x) \sin((k_0 + \frac{\delta}{2})x) + \notag\\
	\epsilon \, \cos((2 \, k_0 + \delta) x) \, \big(a(x) \cos((k_0 + \frac{\delta}{2})x) + b(x) \sin((k_0 + \frac{\delta}{2})x)	\big)
\end{gather}

\begin{gather}
	\epsilon \, \cos((2 \, k_0 + \delta) x) \, \big(a(x) \cos((k_0 + \frac{\delta}{2})x) + b(x) \sin((k_0 + \frac{\delta}{2})x)	\big) \notag\\
	= \notag\\
	\epsilon \, a(x) \cos((2 \, k_0 + \delta) x) \,  \cos((k_0 + \frac{\delta}{2})x) + \notag\\
	\epsilon \, b(x) \cos((2 \, k_0 + \delta) x) \,  \sin((k_0 + \frac{\delta}{2})x)  \notag\\
	=  \notag\\
	\dfrac{\epsilon \, a(x)}{2} \big( \xcancel{\cos((3 \, k_0 + \frac{3 \delta}{2}) x)}  + \cos(( k_0 + \frac{ \delta}{2}) x)  \big) + \notag\\
	\dfrac{\epsilon \, b(x)}{2} \big( \xcancel{\sin((3 \, k_0 + \frac{3 \delta}{2}) x)}  - \sin(( k_0 + \frac{ \delta}{2}) x)  \big)
	= \notag\\
	\dfrac{\epsilon \, a(x)}{2} \big(  \cos(( k_0 + \frac{ \delta}{2}) x)  \big) - \notag\\
	\dfrac{\epsilon \, b(x)}{2} \big( \sin(( k_0 + \frac{ \delta}{2}) x)  \big)
\end{gather}

\begin{gather}
	(1 + \epsilon \, \cos((2 \, k_0 + \delta) x)) \, \textbf{n}_{\omega} \notag\\
	= \notag\\
	\big(a(x) \cos((k_0 + \frac{\delta}{2})x) + b(x) \sin((k_0 + \frac{\delta}{2})x)	\big) +\notag\\
	\dfrac{\epsilon \, a(x)}{2} \big(  \cos(( k_0 + \frac{ \delta}{2}) x)  \big) - \notag\\
	\dfrac{\epsilon \, b(x)}{2} \big( \sin(( k_0 + \frac{ \delta}{2}) x)  \big) \notag\\
	= \notag\\
	a(x) \big( 1 + \dfrac{\epsilon}{2}\big) \cos((k_0 + \frac{\delta}{2})x) + b(x) \big( 1 - \dfrac{\epsilon}{2}\big) \sin((k_0 + \frac{\delta}{2})x)	
\end{gather}

Here we neglect the high order oscillations with the wavenumber of  $3k_0$ and consider that both the coefficients in front of   $\cos((k_0 + \frac{\delta}{2})x)$ and   $\sin((k_0 + \frac{\delta}{2})x)$ must be zero to satisfy the differential equation identically.

\begin{gather}
	\big(  -2 \dfrac{da(x)}{dx} (k_0 + \frac{\delta}{2}) - b(x) (k_0 + \frac{\delta}{2})^2 \big) \sin((k_0 + \frac{\delta}{2})x) - \notag\\
	\big(  2 \dfrac{db(x)}{dx} (k_0 + \frac{\delta}{2}) - a(x) (k_0 + \frac{\delta}{2})^2 \big) \cos((k_0 + \frac{\delta}{2})x) \notag\\
	= \notag\\
	-k_0^2 \Big( a(x) \big( 1 + \dfrac{\epsilon}{2}\big) \cos((k_0 + \frac{\delta}{2})x) + b(x) \big( 1 - \dfrac{\epsilon}{2}\big) \sin((k_0 + \frac{\delta}{2})x) \Big)
\end{gather}

We get the system of two differential equation with respect  to $a(x)$ and $b(x)$.

\begin{gather}
	-2 \dfrac{da(x)}{dx} (k_0 + \frac{\delta}{2}) - b(x) (k_0 + \frac{\delta}{2})^2 + k_0^2 b(x) \big( 1 - \dfrac{\epsilon}{2}\big)  = 0 \notag\\
	2 \dfrac{db(x)}{dx} (k_0 + \frac{\delta}{2}) - a(x) (k_0 + \frac{\delta}{2})^2  + k_0^2 a(x) \big( 1 + \dfrac{\epsilon}{2}\big)  = 0
\end{gather}

\end{document}